# Big Geo Data Surface Approximation using Radial Basis Functions: A Comparative Study

Zuzana Majdisova[a,*], Vaclav Skala[a]

[a]*Department of Computer Science and Engineering, Faculty of Applied Sciences, University of West Bohemia, Univerzitní 8, CZ 30614 Plzeň, Czech Republic*

## Abstract

Approximation of scattered data is often a task in many engineering problems. The Radial Basis Function (RBF) approximation is appropriate for big scattered datasets in $n$−dimensional space. It is a non-separable approximation, as it is based on the distance between two points. This method leads to the solution of an overdetermined linear system of equations.

In this paper the RBF approximation methods are briefly described, a new approach to the RBF approximation of big datasets is presented, and a comparison for different Compactly Supported RBFs (CS-RBFs) is made with respect to the accuracy of the computation. The proposed approach uses symmetry of a matrix, partitioning the matrix into blocks and data structures for storage of the sparse matrix. The experiments are performed for synthetic and real datasets.

*Keywords:*
Radial basis functions, CS-RBF, Approximation, Wendland's RBF, Big data, Point clouds

## 1. Introduction

Interpolation and approximation are the most frequent operations used in computational techniques. Several techniques have been developed for data interpolation or approximation, but they usually require an ordered dataset, e.g. rectangular mesh, structured mesh, unstructured mesh, etc. However, in many engineering problems, data are not ordered and they are scattered in $n$−dimensional space, in general. Usually, in technical applications the conversion of a scattered dataset to a semi-regular grid is performed using some tessellation techniques. However, this approach is quite prohibitive for the case of $n$−dimensional data due to the computational cost.

Interesting techniques are based on the Radial Basis Function (RBF) method, which was originally introduced by Hardy (1971), Hardy (1990). A good introduction to RBFs is given by Buhmann (2003). RBF techniques are widely used across many fields solving technical and non-technical problems, e.g. surface reconstruction (Carr et al. (2001), Turk and O'Brien (2002)), data visualization (Pepper et al. (2014)) and pattern recognition. It is an effective tool for solving partial differential equations (Hon et al. (2015), Li et al. (2013)). The RBF techniques are really meshless and are based on collocation in a set of scattered nodes. These methods are independent with respect to the dimension of the space. The computational cost of the RBF approximation increases nonlinearly (almost cubic) with the number of points in the given dataset and linearly with the dimensionality of the data. Of course, there are other meshless techniques such as discrete smooth interpolation (DSI) (Mallet (1989)), kriging (Royer and Vieira (1984), Ma et al. (2014), Cressie (2015)), which is based on statistical models that include autocorrelation, etc.

The radial basis functions are divided into two main groups of basis functions: global RBFs and Compactly Supported RBFs (CS−RBFs) (Wendland (2006)). In this paper, we will mainly focus on CS-RBFs. Fitting scattered data with CS−RBFs leads to a simpler and faster computation, because the system of linear equations has a sparse matrix. However, an approximation using CS−RBFs is sensitive to the density of the approximated scattered data and to the choice of a shape parameter. Global RBFs are useful in repairing incomplete datasets and they are insensitive to the density of scattered data. However, global RBFs lead to a linear system of equations with a dense matrix and therefore they have high computational and memory costs. Typical global RBFs are Gauss $\phi(r) = e^{-(\alpha r)^2}$, inverse quadratic $(1 + (\alpha r)^2)^{-1}$ and inverse multiquadric $(1 + (\alpha r)^2)^{-1/2}$, where $\alpha$ is shape parameter which defines behavior of function. These RBFs are monotonically decreased with increasing radius $r$, strictly positive definite, infinitely differentiable and convergent to zero. Other global RBF is multiquadric $\sqrt{1 + (\alpha r)^2}$ which is monotonically increased with increasing radius $r$, infinitely differentiable and divergent as radius increases. The last popular global RBF is thin plate spline (TPS) $r^2 \log(r)$ which is shape parameter free and divergent as radius increases. TPS has a singularity at the origin which is removable for the function and its first derivative but this singularity is not removable for the second derivative of TPS.

For the processing of scattered data we can use the RBF interpolation or the RBF approximation. The unknown function sampled at given points $\{x_i\}_1^N$ by values $\{h_i\}_1^N$ can be determined using the RBF interpolation, e.g. presented by Skala (2015), as:

---
[*]Corresponding author
*Email address:* majdisz@kiv.zcu.cz (Zuzana Majdisova)
*URL:* www.vaclavskala.eu (Vaclav Skala)

$$f(\mathbf{x}) = \sum_{j=1}^{N} c_j \phi(r_j) = \sum_{j=1}^{N} c_j \phi(\|\mathbf{x} - \mathbf{x}_j\|), \quad (1)$$

where the interpolating function $f(\mathbf{x})$ is represented as a sum of $N$ RBFs, each centered at a different data point $\mathbf{x}_j$ and weighted by an appropriate weight $c_j$ which has to be determined. This leads to a solution of linear system of equations:

$$A\mathbf{c} = \mathbf{h}, \quad (2)$$

where the matrix $\mathbf{A} = \{A_{ij}\} = \{\phi(\|\mathbf{x}_i - \mathbf{x}_j\|)\}$ is $N \times N$ symmetric square matrix, the vector $\mathbf{c} = (c_1, \ldots, c_N)^T$ is the vector of unknown weights and $\mathbf{h} = (h_1, \ldots, h_N)^T$ is a vector of values in the given points. The disadvantage of RBF interpolation is the large and usually ill-conditioned matrix of the linear system of equations. Note that the one of the possible solution of ill-condition problems based on modified orthogonal least squares is described in Chen and Li (2012). Moreover, in the case of an oversampled dataset or intended reduction, we want to reduce the given problem, i.e. reduce the number of weights and used basis functions, and preserve good precision of the approximated solution. The approach which includes such a reduction is called the RBF approximation. In the following section, the approach recently introduced in Skala (2013) will be described in detail. This approach requires less memory and offers higher speed of computation than the method using Lagrange multipliers (Fasshauer (2007)). Further, a new approach to RBF approximation of large datasets is presented in Section 5. This approach uses symmetry of a matrix, partitioning the matrix into blocks and data structures for storage of the sparse matrix (see Section 4).

## 2. RBF Approximation

For simplicity, we assume that we have an unordered dataset $\{\mathbf{x}_i\}_1^N \in E^2$. However, this approach is generally applicable for $n$-dimensional space. Further, each point $\mathbf{x}_i$ from the dataset is associated with a vector $\mathbf{h}_i \in E^p$ of the given values, where $p$ is the dimension of the vector, or a scalar value, i.e. $h_i \in E^1$. For an explanation of the RBF approximation, let us consider the case when each point $\mathbf{x}_i$ is associated with a scalar value $h_i$, e.g. a $2^1/_2 D$ surface. Let us introduce a set of new reference points (knots of RBF) $\{\boldsymbol{\xi}_j\}_1^M$, see Figure 1.

These reference points may not necessarily be in a uniform grid. A good placement of the reference points improves the approximation of the underlying data. For example, when a terrain is approximated, placement along features such as break lines leads to better approximation results. The number of reference points $\boldsymbol{\xi}_j$ is $M$, where $M \ll N$. The RBF approximation is based on the distance computation between the given point $\mathbf{x}_i$ and the reference point $\boldsymbol{\xi}_j$.

The approximated value is determined as (see Skala (2013)):

$$f(\mathbf{x}) = \sum_{j=1}^{M} c_j \phi(r_j) = \sum_{j=1}^{M} c_j \phi(\|\mathbf{x} - \boldsymbol{\xi}_j\|), \quad (3)$$

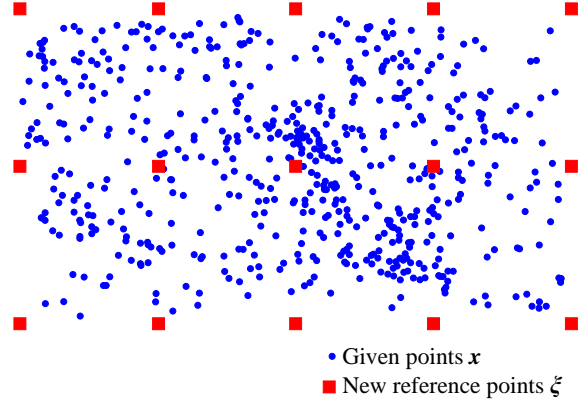

- Given points $\mathbf{x}$
- New reference points $\boldsymbol{\xi}$

Figure 1: The RBF approximation and reduction of points. Note that the reference points (knots) can be distributed arbitrarily.

where $\phi(r_j)$ is an RBF centered at point $\boldsymbol{\xi}_j$ and the approximating function $f(\mathbf{x})$ is represented as a sum of these RBFs, each associated with a different reference point $\boldsymbol{\xi}_j$, and weighted by a coefficient $c_j$ which has to be determined.

When inserting all data points $\mathbf{x}_i$, with $i = 1, \ldots, N$, into (3), we get an overdetermined linear system of equations.

$$h_i = f(\mathbf{x}_i) = \sum_{j=1}^{M} c_j \phi(\|\mathbf{x}_i - \boldsymbol{\xi}_j\|) = \sum_{j=1}^{M} c_j \phi_{i,j} \quad i = 1, \ldots, N \quad (4)$$

The linear system of equations (4) can be represented in a matrix form as:

$$A\mathbf{c} = \mathbf{h}, \quad (5)$$

where $A_{ij} = \phi(\|\mathbf{x}_i - \boldsymbol{\xi}_j\|)$ is the entry of the matrix in the $i$-th row and $j$-th column, the number of rows is $N \gg M$, $M$ is the number of unknown weights $\mathbf{c} = (c_1, \ldots, c_M)^T$, i.e. a number of reference points, and $\mathbf{h} = (h_1, \ldots, h_N)^T$ is a vector of values in the given points. The presented system is overdetermined, i.e. the number of equations $N$ is higher than the number of variables $M$. This linear system of equations can be solved by the least squares method (LSE) as $A^T A \mathbf{c} = A^T \mathbf{h}$.

## 3. RBF Approximation with Polynomial Reproduction

The method which was described in Section 2 can have problems with stability and solvability. Therefore, the RBF approximant (3) is usually extended by a polynomial function $P_k(\mathbf{x})$ of the degree $k$. This approach was introduced in Majdisova and Skala (2016).

The approximated value $f(\mathbf{x})$ is determined as:

$$f(\mathbf{x}) = \sum_{j=1}^{M} c_j \phi(\|\mathbf{x} - \boldsymbol{\xi}_j\|) + P_k(\mathbf{x}), \quad (6)$$

where $\boldsymbol{\xi}_j$ are reference points specified by a user. The approximating function $f(\mathbf{x})$ is represented as a sum of $M$ RBFs, each associated with a different reference point $\boldsymbol{\xi}_j$, and weighted by an appropriate coefficient $c_j$, and $P_k(x)$ is a polynomial function of degree $k$. It should be noted that the polynomial function affects only global behavior of the approximated dataset. In practice, a linear polynomial $P_1(x)$:

$$P_1(x) = a^T x + a_0 \quad (7)$$

is used (e.g. $P_1(x) = a_1 x + a_2 y + a_0$ for $x \in E^2$). Geometrically, the coefficient $a_0$ determines the "vertical" placement of the hyperplane and the expression $a^T x$ represents the inclination of the hyperplane.

Thus, the following overdetermined linear system of equations is obtained:

$$h_i = f(x_i) = \sum_{j=1}^{M} c_j \phi(\|x_i - \xi_j\|) + a^T x + a_0$$
$$= \sum_{j=1}^{M} c_j \phi_{i,j} + a^T x + a_0 \quad i = 1, \ldots, N. \quad (8)$$

The linear system of equations (8) can be represented in a matrix form as:

$$Ac + Pk = h, \quad (9)$$

where $A_{ij} = \phi(\|x_i - \xi_j\|)$ is the entry of the matrix in the $i$-th row and $j$-th column, $c = (c_1, \ldots, c_M)^T$ is the vector of unknown weights, $P_i = (x_i^T, 1)$ is the vector of basis functions of linear polynomial at point $x_i$, $k = (a^T, a_0)^T$ is the vector of the coefficient for the linear polynomial and $h = (h_1, \ldots, h_N)^T$ is the vector of values in the given points. The presented linear system of equations can be solved by the minimization of the square of error, which leads to a system of linear equations:

$$\begin{pmatrix} A^T A & A^T P \\ P^T A & P^T P \end{pmatrix} \begin{pmatrix} c \\ k \end{pmatrix} = \begin{pmatrix} A^T h \\ P^T h \end{pmatrix}. \quad (10)$$

Finally, it should be noted that the polynomial of degree $k > 1$ can be used in general. However, in this case, it is necessary be careful because the polynomial of higher degree in combination with a large range of data might cause numerical problems. This is due to the fact that the elements of sub-matrix $P^T P$ in relation (10) contain much larger values than elements of sub-matrix $A^T A$ in the same relation.

## 4. Data Structures for Storage of the Sparse Matrix

If the CS-RBFs are used, the matrix of the linear system of equations is sparse. Therefore, the most important part of each approximation using CS-RBFs is a data structure used to store the approximation matrix. There are a number of existing sparse matrix representations, e.g. Bell and Garland (2009), Šimeček (2009), each with different computational characteristics, storage requirements and methods of accessing and manipulating entries of the matrix. The main difference among existing storage formats is the sparsity pattern, or the structure of the nonzero elements, for which they are best suited. For our purpose, the coordinate format is used, which is briefly described in the following.

The coordinate (COO) format is the simplest storage scheme. The sparse matrix is represented by three arrays: data, where the $N_{NZ}$ nonzero values are stored, row, where the row index of each nonzero element is kept, and col, where the column indices of the nonzero values are stored.

Example of the COO format for matrix $Q$:

$$Q = \begin{pmatrix} 1 & 0 & 6 & 0 & 0 \\ 9 & 2 & 0 & 7 & 0 \\ 0 & 1 & 3 & 0 & 8 \\ 4 & 0 & 2 & 4 & 0 \\ 0 & 5 & 0 & 0 & 0 \end{pmatrix}$$

```
row  = [ 0  0  1  1  1  2  2  2  3  3  3  4 ]
col  = [ 0  2  0  1  3  1  2  4  0  2  3  1 ]
data = [ 1  6  9  2  7  1  3  8  4  2  4  5 ]
```

So, if the COO format is used for representation of matrix $Q$ (in form as described above) and the equation $y = Qx$, where $x$ is vector of the given values, has been solved, the following pseudocode is used for calculation:

$\forall i = 0, \ldots, N : y_{[i]} = 0$
**for** $i = 0, \ldots, N_{NZ} - 1$ **do**
  $y_{\text{row}[i]} = y_{\text{row}[i]} + \text{data}_{[i]} \cdot x_{\text{col}[i]}$

Note that vector of given values has form $x = [x_0, x_1, \ldots, x_M]$, where $M$ is number of columns of matrix $Q$, and the resulting vector is $y = [y_0, y_1, \ldots, y_N]$, where $N$ is number of rows of matrix $Q$.

The benefit of the COO format is its generality, i.e. an arbitrary sparse matrix can be represented by the COO format and the required storage is always proportional to the number of nonzero values.

The disadvantage of the COO format is that both row and column indices are stored explicitly, which reduces the efficiency of memory transactions (e.g. read operations).

## 5. RBF Approximation for Large Data

In practice, real datasets contain a large number of points, which results in high memory requirements for storing the matrix $A$ of the overdetermined linear system of equations (5). Unfortunately, we do not have an unlimited capacity of RAM memory; therefore, calculation of unknown weights $c_j$ for RBF approximation would be prohibitively computationally expensive due to memory swapping, etc. In this section, a proposed solution to this problem is described.

In Section 2, it was mentioned that an overdetermined system of equations can be solved by the least squares method. For this method the square $M \times M$ matrix:

$$B = A^T A \quad (11)$$

is to be determined. Advantages for computation of the matrix $B$ are that it is a symmetric matrix and, moreover, only two vectors of length $N$ are needed for determination of one entry, i.e.:

$$b_{ij} = \sum_{k=1}^{N} \phi_{ki} \cdot \phi_{kj}, \quad (12)$$

where $b_{ij}$ is the entry of the matrix $\boldsymbol{B}$ in the $i$–th row and $j$–th column.

To save memory requirements and to prevent data bus (PCI) overloading, block operations with matrices are used. Based on the above properties of the matrix $\boldsymbol{B}$, only the upper triangle of this matrix is computed. Moreover, the matrix $\boldsymbol{B}$ is partitioned into $M_B \times M_B$ blocks, see Figure 2, and the calculation is performed sequentially for each block:

$$\boldsymbol{B}_{kl} = (\boldsymbol{A}_{*,k})^T (\boldsymbol{A}_{*,l})$$
$$k = 1, \ldots, \left\lceil \frac{M}{M_B} \right\rceil, \qquad l = k, \ldots, \left\lceil \frac{M}{M_B} \right\rceil, \qquad (13)$$

where $\boldsymbol{B}_{kl}$ is a sub-matrix in the $k$–th row and $l$–th column, the index $*$ denotes that the sub-matrix $\boldsymbol{A}_{*,k}$ contains all values in the appropriate block of columns (given by the index $k$) of the original matrix $\boldsymbol{A}$, i.e. $\boldsymbol{A}_{*,k}$ is defined as:

$$\boldsymbol{A}_{*,k} = \begin{pmatrix} \phi_{1,(k-1)\cdot M_B+1} & \cdots & \phi_{1,\min\{k\cdot M_B,M\}} \\ \vdots & \ddots & \vdots \\ \phi_{i,(k-1)\cdot M_B+1} & \cdots & \phi_{i,\min\{k\cdot M_B,M\}} \\ \vdots & \ddots & \vdots \\ \phi_{N,(k-1)\cdot M_B+1} & \cdots & \phi_{N,\min\{k\cdot M_B,M\}} \end{pmatrix}, \qquad (14)$$

where the size of this matrix is $N \times M_B$ except of the last block and the index $k$ denotes the $k$–th block of columns. This enables the computation of big datasets on hardware systems with limited main memory.

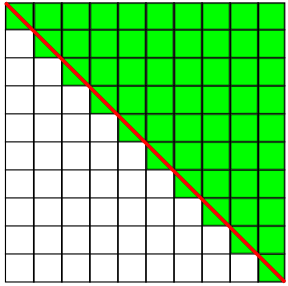

Figure 2: $M \times M$ square matrix which is partitioned into $M_B \times M_B$ blocks. The color red is used to denote the main diagonal of the matrix and illustrates the symmetry of the matrix. The color green is used to denote the blocks which must be computed.

The size of block $M_B$ is chosen so that swapping of memory (RAM) does not occur during the computation, i.e.:

$$(M^2 + 2 \cdot M_B \cdot N) \cdot prec < \text{size of RAM [B]}, \qquad (15)$$

where $prec$ is the size of the data type in bytes. Note that this relation is valid when the matrix $\boldsymbol{A}$ of the overdetermined linear system of equations is dense. If CS-RBFs are used for RBF approximation and the matrix $\boldsymbol{A}$ of the overdetermined linear system of the equation is stored using special data structures, see Section 4, then the optimal size of block $M_B$ is much larger than given in relation (15). For this case, the optimal size of block $M_B$ should satisfy:

$$(M^2 + 2 \cdot N_{NZ}) \cdot prec < \text{size of RAM [B]}, \qquad (16)$$

where $N_{NZ}$ is the maximum number of non-zero elements in sub-matrices $\boldsymbol{A}_{*,k}$, $k = 1, \ldots, \left\lceil \frac{M}{M_B} \right\rceil$. Naturally, it is obvious that the size of the block should be selected as the largest possible value satisfying (16).

Moreover, note that the elements in sub-matrices $\boldsymbol{A}_{*,k}$ are zero for far away points, when CS-RBFs are used. Therefore, we do not want to compute the elements for all pairs of points, so the $kd$-tree (A.2 in Fasshauer (2007)) is used for computing the sub-matrices $\boldsymbol{A}_{*,k}$. Algorithm for determination of the sparse sub-matrix $\boldsymbol{A}_{*,k}$ is described in Algorithm 1.

---

**Algorithm 1** Determination of the sub-matrix $\boldsymbol{A}_{*,k}$ when CS-RBFs are used. Note that the order of the elements in the triplet (row 5) is {row index, col index, value}

---

**Input:** given points $\{\boldsymbol{x}_i\}_1^N$, reference points $\{\boldsymbol{\xi}_i\}_{(k-1)\cdot M_B}^{\min\{k\cdot M_B,M\}}$, shape parameter $\alpha$, CS-RBF $\phi$
**Output:** sub-matrix $\boldsymbol{A}_{*,k}$ in COO format, i.e. return three arrays `row`, `col`, `data`
1: Build a $kd$-tree for the given points $\{\boldsymbol{x}_i\}_1^N$
2: **for each** reference point $\boldsymbol{\xi}_j$ **do**
3:     Query the $kd$-tree for points $\{\boldsymbol{x}_q\}$ such that $\|\boldsymbol{x}_q - \boldsymbol{\xi}_j\| < \frac{1}{\alpha}$
4:     **for each** point in a support radius $\boldsymbol{x}_q$ **do**
5:         Add triplet $\{q, j, \phi(\|\boldsymbol{x}_q - \boldsymbol{\xi}_j\|)\}$ to COO format

---

In general, the mentioned approach could be used in combination with massive parallel computing on GPU, but the calculation would have to be done in single precision to exploit the full potential of GPU. However, in this case, problems with numerical stability and solvability of the RBF approximation can be expected.

Finally, note that it is possible to modify this approach easily for the RBF approximation with a polynomial reproduction, see Section 3.

## 6. Experimental Results

The presented RBF approximation method was tested on synthetic and real data. The implementation was performed in Matlab. Experimental results for one synthetic and two real datasets follow.

The synthetic dataset has a Halton distribution (A.1 in Fasshauer (2007)) of points and each point is associated with a value from Franke's function (Franke (1979)):

$$f(\boldsymbol{x}) = f_1(\boldsymbol{x}) + f_2(\boldsymbol{x}) + f_3(\boldsymbol{x}) - f_4(\boldsymbol{x}),$$
$$f_1(\boldsymbol{x}) = 0.75 \cdot \exp\left(-\frac{(9x_1 - 2)^2}{4} - \frac{(9x_2 - 2)^2}{4}\right),$$
$$f_2(\boldsymbol{x}) = 0.75 \cdot \exp\left(-\frac{(9x_1 + 1)^2}{49} - \frac{(9x_2 + 1)^2}{10}\right), \qquad (17)$$
$$f_3(\boldsymbol{x}) = 0.50 \cdot \exp\left(-\frac{(9x_1 - 7)^2}{4} - \frac{(9x_2 - 3)^2}{4}\right),$$
$$f_4(\boldsymbol{x}) = 0.20 \cdot \exp\left(-(9x_1 - 4)^2 - (9x_2 - 7)^2\right),$$

where $\boldsymbol{x} = (x_1, x_2)$ is a point for which the associated value has been computed. This function is shown in Figure 3.

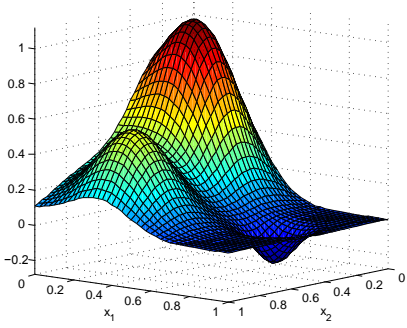

Figure 3: Franke's function defined as (17).

Table 1: Overview information for the tested datasets. The Axis-Aligned Bounding Boxes (AABBs) of the tested datasets have a size width × length × relief, i.e. $x_{range} \times y_{range} \times z_{range}$. Note that one foot [ft] corresponds to 0.3048 meter [m].

|  | Synth. | Serpent Mound | St. Helens |
|---|---|---|---|
| **number of pts.** | 1089 | 3,265,110 | 6,743,176 |
| **number of ref. pts.** | 81 | 10,000 | 10,000 |
| **relief [ft]** | 1.238 | 48.70 | 5138.69 |
| **width [ft]** | 1.000 | 1,085.12 | 26,232.37 |
| **length [ft]** | 1.000 | 2,698.96 | 35,992.69 |

The first real dataset was obtained from LiDAR data of Mount Saint Helens in Skamania County, Washington[1], see Figure 4 (left). The second real dataset is LiDAR data of the Serpent Mound in Adams County, Ohio[1], see Figure 4 (right). Each point of these datasets is associated with its elevation.

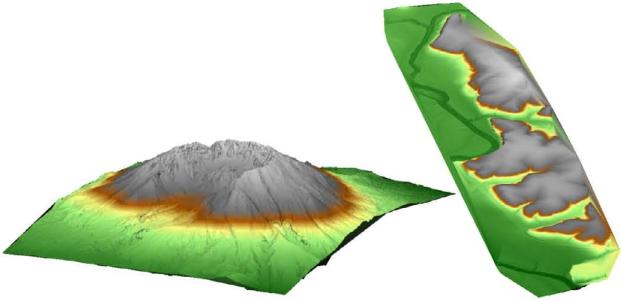

Figure 4: Original datasets: Mount Saint Helens (left); Serpent Mound (right).

Moreover, as a first step, the real datasets are translated so that their estimated center of gravity corresponds to the origin of the coordinate system. This step is used due to the limitation of the influence of dataset placement in space. The set of reference points is a subset of the given dataset, for which we determine the RBF approximation. In addition, reference points are uniformly distributed within a given area. Table 1 gives an overview of the used datasets.

Because the global RBFs affect the entire domain of given datasets, which is usually undesirable behavior, the CS-RBFs have been used for the presented experiments. All CS-RBFs from the catalog of RBFs in Fasshauer (2007) (see D.2.7) have been used for the experiments. Depending on the quality, the obtained results are divided into three groups. The results are presented for a representative of each group, see Table 2. Note that the notation $(1 - \alpha r)^q_+$ means:

$$(1 - \alpha r)^q_+ = \begin{cases} (1 - \alpha r)^q & \text{if} \quad 0 \leq \alpha r \leq 1, \\ 0 & \text{if} \quad \alpha r > 1 \end{cases}, \quad (18)$$

where $r$ is the variable which denotes the distance of the given point from the appropriate reference point and $\alpha$ is a shape parameter. The shape parameters $\alpha$ for the used CS-RBFs were determined experimentally with regard to the quality of approximation and they are presented in Table 3. Some papers have also been published on choosing the optimal shape parameter $\alpha$, e.g. Franke (1982), Rippa (1999), Fasshauer and Zhang (2007), Scheuerer (2011). Note that the value of the shape parameter $\alpha$ is inversely proportional to the width, length, and number of points of the datasets.

[1] http://www.liblas.org/samples/

Table 2: Used Wendland's CS-RBFs $\phi_{d,s}$. Wendland's functions are univariate polynomial of degree $\lfloor d/2 \rfloor + 3s + 1$, they are always positive definite up to a maximal space dimension $d$ and their smoothness is $C^{2s}$. For more details see Chapter 11.2 in Fasshauer (2007).

| CS-RBF | $\phi(r)$ |
|---|---|
| $\phi_{3,0}$ | $(1 - \alpha r)^2_+$ |
| $\phi_{3,1}$ | $(1 - \alpha r)^4_+ (4\alpha r + 1)$ |
| $\phi_{3,3}$ | $(1 - \alpha r)^8_+ (32(\alpha r)^3 + 25(\alpha r)^2 + 8\alpha r + 1)$ |

Table 3: Experimentally determined shape parameters $\alpha$ for the used CS-RBFs

| CS-RBF | shape parameter | | |
|---|---|---|---|
|  | Synthetic | Serpent Mound | St. Helens |
| Wendland's $\phi_{3,0}$ | $\alpha = 0.707$ | $\alpha = 0.01$ | $\alpha = 0.0005$ |
| Wendland's $\phi_{3,1}$ | $\alpha = 0.500$ | $\alpha = 0.01$ | $\alpha = 0.0007$ |
| Wendland's $\phi_{3,3}$ | $\alpha = 0.250$ | $\alpha = 0.01$ | $\alpha = 0.0005$ |

Figure 5 presents the approximations of the synthetic dataset without polynomial reproduction for all CS−RBFs. In this figure, the surfaces are false-colored by the magnitude of the error. The error is defined as the absolute value of the difference between Franke's function (17) and approximated function. It can be seen that for the synthetic dataset, the RBF approximation with Wendland's $\phi_{3,3}$ basis function returns the best result in terms of the error. On the contrary, the worst result is obtained for the RBF approximation with Wendland's $\phi_{3,0}$ basis function. Table 4 shows three different error measures of the datasets depending on the chosen basis functions: mean absolute error, deviation and mean relative error. These error measures are performed for approximation without polynomial reproduction and for approximation with linear polynomial reproduction. It can be seen that the RBF approximation with linear polynomial reproduction produces slightly better results than the RBF approximation without reproduction in terms of the error, but this improvement seems to be insignificant.

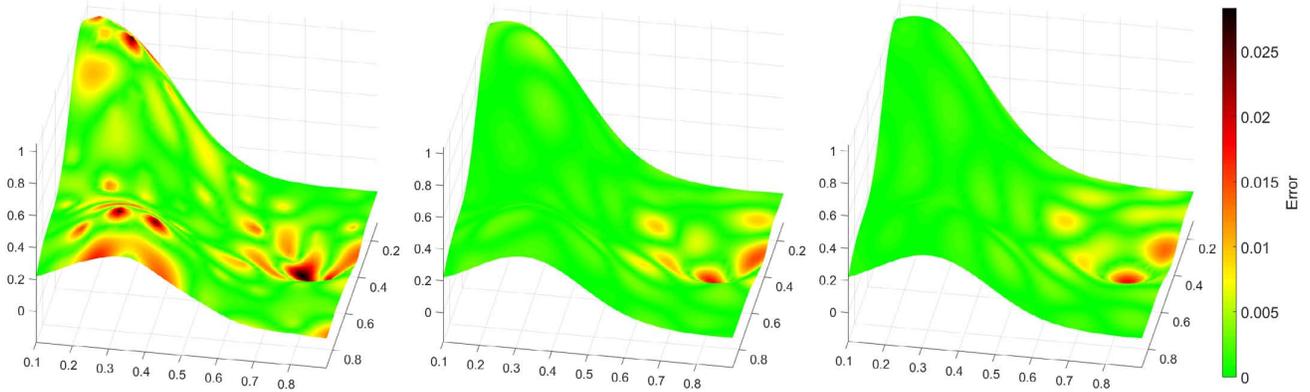

Figure 5: Results for synthetic dataset false-colored by magnitude of absolute error: Wendland's RBF $\phi_{3,0}$, $\alpha = 0.707$ (left); Wendland's RBF $\phi_{3,1}$, $\alpha = 0.500$ (center) and Wendland's RBF $\phi_{3,3}$, $\alpha = 0.250$ (right).

The RBF approximation for the real datasets was solved using "block-wise" approach described above. Approximations of Mount Saint Helens dataset without polynomial reproduction for all CS-RBFs are shown in Figure 6a. It illustrates the magnitude of error at each point of the original point cloud. Moreover, the detail of a crater is shown for each approximation. It can be seen that the RBF approximation with Wendland's $\phi_{3,3}$ basis function returns the best results in terms of the error. On the contrary, the worst result is obtained for the RBF approximation with Wendland's $\phi_{3,0}$ basis function again. For this approximation, sharp peaks are formed. It is most evident around the rim of a crater. Also for the Mount Saint Helens dataset, the three error measures of the computed elevation for all used CS-RBFs and for both types of RBF approximation (i.e. approximation without polynomial reproduction and approximation with linear polynomial reproduction) are presented

Table 4: The RBF approximation error and density of least square matrix for the tested datasets and different radial basis functions. Note that density of least square matrix expresses percentage of non-zero elements in matrix and that one foot [ft] corresponds to 0.3048 meter [m].

| **Phenomenon** | without polynomial | | | with linear polynomial | | |
|---|---|---|---|---|---|---|
| | Wendland's | | | Wendland's | | |
| | $\phi_{3,0}$ | $\phi_{3,1}$ | $\phi_{3,3}$ | $\phi_{3,0}$ | $\phi_{3,1}$ | $\phi_{3,3}$ |
| **Synthetic data** | | | | | | |
| **mean absolute error [ft]** | 0.0041 | 0.0021 | 0.0019 | 0.0040 | 0.0019 | 0.0019 |
| **deviation of error [ft]** | 1.92E-5 | 6.06E-6 | 5.25E-6 | 1.90E-5 | 5.45E-6 | 5.12E-6 |
| **mean relative error [%]** | 0.0151 | 0.0076 | 0.0072 | 0.0150 | 0.0070 | 0.0072 |
| **Serpent Mound** | | | | | | |
| **mean absolute error [ft]** | 0.173 | 0.141 | 0.130 | 0.164 | 0.139 | 0.129 |
| **deviation of error [ft]** | 0.072 | 0.047 | 0.037 | 0.068 | 0.047 | 0.037 |
| **mean relative error [%]** | 0.015 | 0.012 | 0.011 | 0.014 | 0.012 | 0.011 |
| **density of LSE matrix [%]** | 8.413 | 8.413 | 8.413 | 8.468 | 8.468 | 8.468 |
| **Mount St. Helens** | | | | | | |
| **mean absolute error [ft]** | 12.568 | 11.589 | 9.881 | 12.129 | 10.935 | 9.773 |
| **deviation of error [ft]** | 188.595 | 165.574 | 100.738 | 159.139 | 122.659 | 98.993 |
| **mean relative error [%]** | 0.013 | 0.012 | 0.010 | 0.012 | 0.011 | 0.010 |
| **density of LSE matrix [%]** | 6.470 | 3.452 | 6.470 | 6.536 | 3.510 | 6.536 |

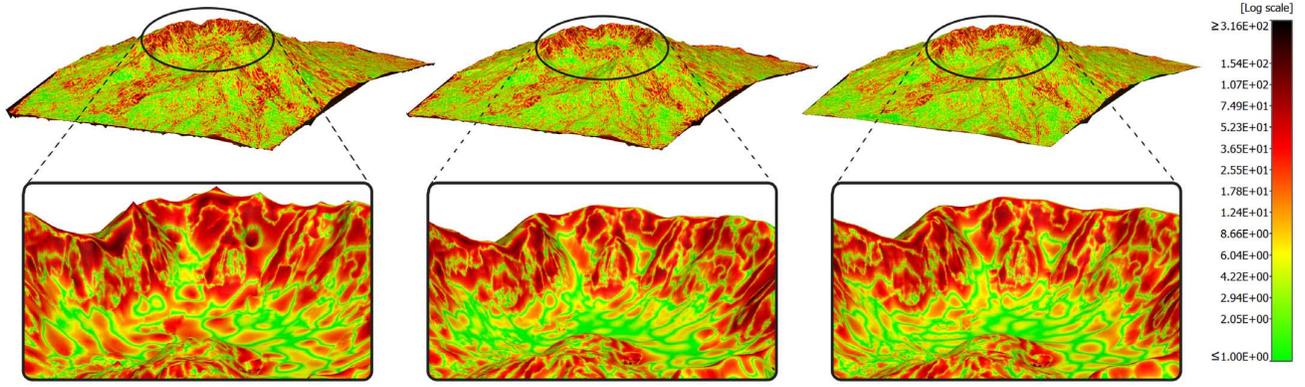

(a) Mount Saint Helens dataset: Wendland's RBF $\phi_{3,0}$, $\alpha = 0.0005$ (left), Wendland's RBF $\phi_{3,1}$, $\alpha = 0.0007$ (center) and Wendland's RBF $\phi_{3,3}$, $\alpha = 0.0005$ (right)

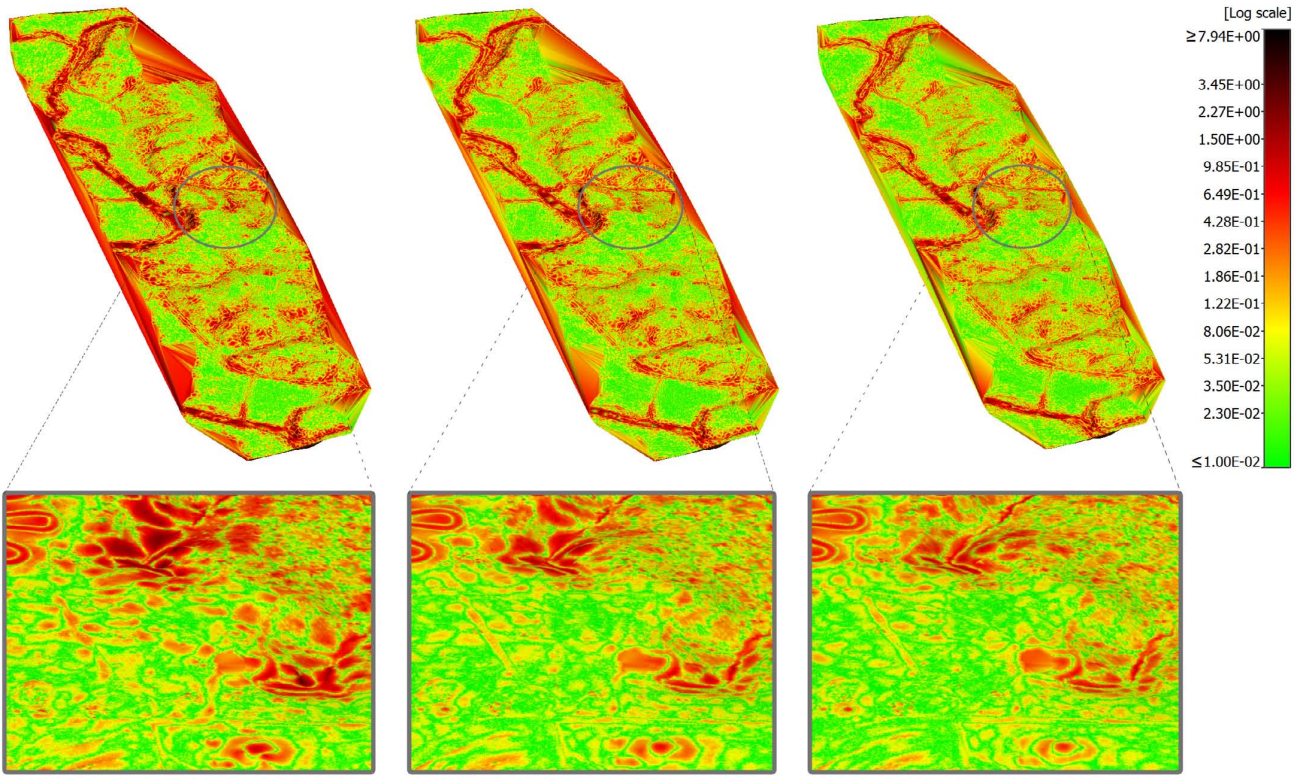

(b) Serpent Mound dataset: Wendland's RBF $\phi_{3,0}$, $\alpha = 0.01$ (left), Wendland's RBF $\phi_{3,1}$, $\alpha = 0.01$ (center) and Wendland's RBF $\phi_{3,3}$, $\alpha = 0.01$ (right) in Table 4. These results confirm the statements above. Further, it can be seen that the RBF approximation with linear reproduction again produces better results than the RBF approximation without reproduction, especially in terms of deviation of error.

Figure 6: Results for the tested real datasets false-colored by magnitude of absolute error.

The last presented experimental results are for the RBF approximation of Serpent Mound without polynomial reproduction and are shown in Figure 6b. It illustrates the magnitude of error at each point of the original point cloud. Moreover, the detail of Serpent Mound is shown for each approximation. It can be seen that the RBF approximation with Wendland's $\phi_{3,3}$ basis function returns a slightly better result than RBF approximation with Wendland's $\phi_{3,1}$ basis function in terms of the error for the Serpent Mound dataset. The RBF approximation with Wendland's $\phi_{3,0}$ basis function returns the worst results. These facts are mainly evident in the details. Further, we can see that the highest errors occur on the boundary of the terrain for all cases. The three error measures of the elevation for all used CS−RBFs and for both types of RBF approximation (i.e. approximation without polynomial reproduction and approximation with linear polynomial reproduction) are pre-sented in Table 4. These results again confirm the statements above. Further, it can be seen that the RBF approximation with linear polynomial reproduction produces slightly better results than the RBF approximation without reproduction in terms of the error, but this improvement is not significant. The mutual comparison of both real datasets in terms of the deviation of error (Table 4) indicates that RBF approximation with linear reproduction returns considerably better results than RBF approximation without polynomial reproduction if the range of associated values is large. Moreover, it should be noted that the degree of smoothness for the tested type of real datasets is lower than degree of smoothness for Wendland's $\phi_{3,1}$ and Wendland's $\phi_{3,3}$ basis functions and, therefore, the comparison of RBF approximation with Wendland's $\phi_{3,1}$ basis function and RBF approximation with Wendland's $\phi_{3,3}$ basis function returns less significant results. The situation is different for comparison of RBF approximation with Wendland's $\phi_{3,0}$ basis function and RBF approximation with Wendland's $\phi_{3,1}$ basis function where the difference is significant. The signed errors for the Serpent Mound dataset and Wendland's $\phi_{3,1}$ basis function are shown in Figure 7. We can see that the signs are different at various locations. Similar results are obtained for the rest of the experiments.

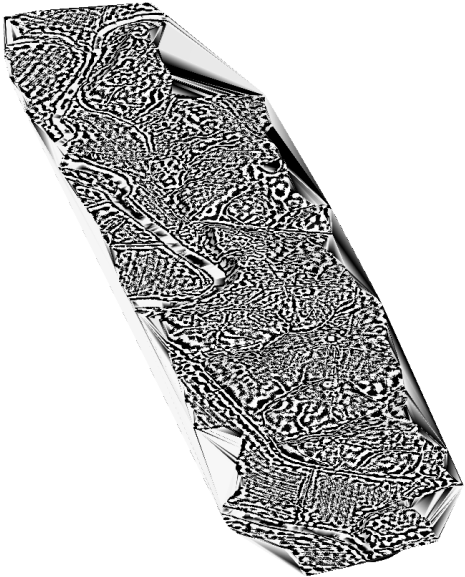

Figure 7: The signed errors for the Serpent Mound dataset and Wendland's RBF $\phi_{3,1}$ with $\alpha = 0.01$: the positive error is colored white and the negative error is colored black.

The implementation of the RBF approximation was performed in MATLAB and tested on a PC with the following configuration:

- CPU: Intel® Core™ i7-4770 (4 × 3.40GHz + hyper-threading),
- memory: 32 GB RAM,
- operation system: Microsoft Windows 7 64 bits.

For the approximation of the Serpent Mound dataset with 10, 000 local Wendland's $\phi_{3,1}$ basis functions with shape parameter $\alpha = 0.01$, the running times for different sizes of blocks were measured. These computational times are presented in Figure 8b. We can see that the the time performance is large for the approximation matrix which is partitioned into small blocks (i.e. smaller than 500 × 500 blocks). This is caused by overhead costs and, moreover, each entry in the matrix $A$ of the overdetermined linear system has to be calculated more times than for larger sizes of block. On the other hand, the running time begins to rise above the permissible limit due to memory swapping for the approximation matrix which is partitioned into larger blocks (i.e. larger than 2500 × 2500 blocks).

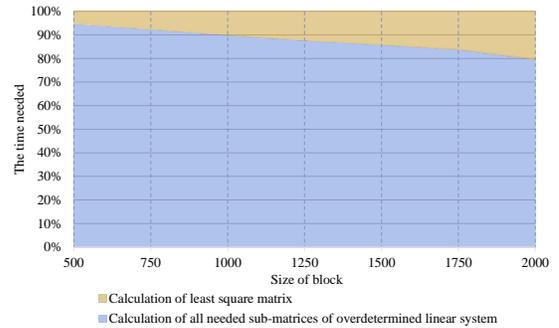

(a) The time needed for calculation of all sub-matrices of the matrix $A$ (blue color) and for determination of least square matrix $A^T A$ (orange color) for the Serpent Mound depending on block size. Note that 100% corresponds to total time of computation.

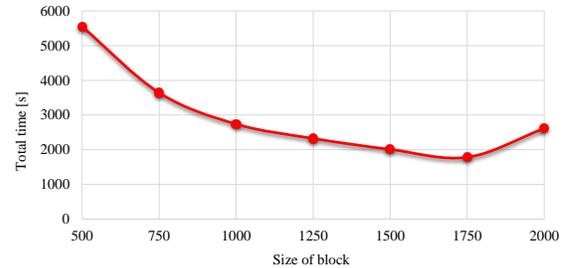

(b) The total time needed for approximation of the Serpent Mound depending on block size.

Figure 8: Time performance for approximation of the Serpent Mound depending on the block size.

The running time for determination of RBF approximation with the mentioned parameters was divided into two steps. The running time needed for calculation of all sub-matrices formed from the matrix $A$ of the original overdetermined linear system of equations by the block-wise approach is determined in the first step. The running time needed for calculation of the least square matrix $A^T A$ and for calculation of the vector of unknown weights is measured in the second step. The comparison of the perceptual time performance of these two steps can be seen in Figure 8a. It can be seen that the most time-consuming part is the first step, in which all needed sub-matrices are calculated (lower part in the graph).

# 7. Conclusion

In this paper two different RBF approximation methods are experimentally verified using one synthetic and two real datasets. The first method is an RBF approximation without polynomial reproduction and the second method is an RBF approximation with linear reproduction. Moreover, a new approach to the RBF approximation of large datasets is presented. The proposed approach uses symmetry of the matrix, partitioning the matrix into blocks and block-wise solving which enables the computation on systems with limited main memory. Because CS-RBFs are used for approximation, data structures for storage of the sparse matrix can be employed; thereby a larger size of blocks can be chosen and the computational costs decrease. The experiments proved that the proposed approach is fully applicable for the RBF approximation for large datasets.

The experiments also showed that, depending on the quality of the results, it is possible to divide the CS-RBFs from the catalog of RBFs (D.2.7 in Fasshauer (2007)) into three groups. The results of the experiments proved that RBF approximation with linear reproduction returns better result than RBF approximation without polynomial reproduction, particularly if the range of associated values is large. The experiments also proved that the RBF methods have problems with the accuracy of calculation on the boundary of an object, which is a well-known property. The presented approach is directly applicable in GIS and geoscience fields.

Future work will be aimed at improving the accuracy at the boundaries, on the computational performance without loss of approximation accuracy and computation of optimal shape parameters. Also, the "moving window" technique will be explored to increase speed of computation.

**Acknowledgments**

The authors would like to thank their colleagues at the University of West Bohemia, Plzeň, for their discussions and suggestions, and the anonymous reviewers for their valuable comments. The research was supported by the National Science Foundation GAČR project GA17-05534S and partially supported by SGS 2016-013 project.**References**